\documentclass[aip,apl,reprint]{revtex4-1}
\usepackage{natbib}
\usepackage{amsmath}    
\usepackage{graphicx}   
\usepackage{verbatim}   
\usepackage{color}     
\usepackage{subfigure}  
\usepackage{ae}
\usepackage{hyperref}   
\usepackage{amsfonts}

\begin{document}

\title{Molecular beam epitaxy of high structural quality Bi$_2$Se$_3$ on lattice matched InP(111) substrates}
\author{S.~Schreyeck}
\author{N.~V.~Tarakina}
\affiliation{Physikalisches Institut, Experimentelle Physik III, Universität Würzburg, Am Hubland, D-97074 Würzburg, Germany}
\author{G.~Karczewski}
\affiliation{Physikalisches Institut, Experimentelle Physik III, Universität Würzburg, Am Hubland, D-97074 Würzburg, Germany}
\affiliation{Institute of Physics, Polish Academy of Sciences, Al. Lotnikov 32/46, 02-668 Warsaw, Poland}
\author{C.~Schumacher}
\author{T.~Borzenko}
\author{C.~Brüne}
\author{H.~Buhmann}
\author{C.~Gould}
\author{K.~Brunner}
\author{L.~W.~Molenkamp}
\affiliation{Physikalisches Institut, Experimentelle Physik III, Universität Würzburg, Am Hubland, D-97074 Würzburg, Germany}
%
%
\begin{abstract}
Epitaxial layers of the topological insulator Bi$_2$Se$_3$ have been grown by molecular beam epitaxy on laterally lattice-matched InP(111)B substrates. High resolution X-ray diffraction shows a significant improvement of Bi$_2$Se$_3$ crystal quality compared to layers deposited on other substrates. The measured full width at half maximum of the rocking curve is $\Delta\omega=13$~arcsec, and the ($\omega-2\theta$) scans exhibit clear layer thickness fringes. Atomic force microscope images show triangular twin domains with sizes increasing with layer thickness. The structural quality of the domains is confirmed on the microscopic level by transmission electron microscopy.
\\
\\
\small \textit{Copyright 2013 American Institute of Physics. This article may be downloaded for personal use only. Any other use requires prior permission of the author and the American Institute of Physics. The following article appeared in Appl. Phys. Lett. 102, 041914 (2013) and may be found at \href{http://link.aip.org/link/?APL/102/041914}{http://link.aip.org/link/?APL/102/041914}.}
\end{abstract}


\maketitle
The V$_2$VI$_3$ compounds (such as Bi$_2$Se$_3$) have been well known as thermoelectric materials for over 50 years. More recently, 3-dimensional (3D) topological insulator (TI) properties have been predicted for a number of these materials and the existence of a topologically protected surface state has been confirmed by angle-resolved photoemission spectroscopy (ARPES).\cite{Zhang2009,Hsieh2009}  While the existence of this surface state is well established, achieving a high structural quality crystal, free of any defects, is very challenging. In an attempt to reduce defect density and to produce thin layers useful for devices, growth of Bi$_2$Se$_3$ layers by molecular beam epitaxy (MBE) has been extensively studied on various substrates, including Si(111),\cite{Zhang2009a,Li2010,He2011,Bansal2011} GaAs(111),\cite{Richardella2010} Al$_2$O$_3$(110),\cite{Tabor2011} SrTiO$_3$(111),\cite{Zhang2011} and CdS(0001).\cite{Kou2011} While the structural quality of these layers has steadily improved, the narrowest rocking curve, an important indicator of structural quality, which has been reported to date, had a full width at half maximum (FWHM) of $\Delta\omega$~$\sim$~360~arcsec,\cite{Richardella2010} i.e., below the quality usually obtained in zincblende semiconductor MBE layers.\cite{Wenisch1996} Here, we make use of nearly lattice matched InP(111)B substrates with low miscut (< 0.2$^\circ$). These have a hexagonal surface lattice with a lattice-mismatch of only 0.16\% to the ab-plane of Bi$_2$Se$_3$.\cite{Seizo1963, Thompson1969} We show that these substrates allow for MBE growth of Bi$_2$Se$_3$ layers with a high crystal quality, which further increases with layer thickness. For a 250~nm thick layer, a high resolution rocking curve with FWHM of $\Delta\omega$~$\sim$~13~arcsec is obtained.

Bi$_2$Se$_3$ has a crystal symmetry belonging to the space group R$\overline{3}$m (D$_{3d}^5$) and has lattice constants of a~=~b~=~4.143~\AA~and c~=~28.636~\AA.\cite{Seizo1963} It consists of so-called quintuple layers (QLs) Se-Bi-Se-Bi-Se with $ABC$ stacking order along the [001] direction. The bonding type within a QL is covalent, in contrast to the weak van der Waals bonding between QLs.\cite{Wiese1960}

Growth of the Bi$_2$Se$_3$ is carried out in a CreaTec UHV MBE system with a base pressure of 10$^{-10}$~mbar. Elemental Bi (6N) and Se (6N) are evaporated from standard effusion cells. The undoped InP(111)B substrates have a miscut to the (111) plane specified to be below 0.2$^\circ$. Atomic force microscope (AFM) measurements of a reference substrate reveal terraces of about 1 $\mu$m length, corresponding to a miscut of 0.02$^\circ$. Before growth, the substrate surface oxide is removed in 50\% hydrofloric acid (HF). The substrate is then transported in an N$_2$ atmosphere and quickly loaded into the MBE system. The growth is monitored by reflection high-energy electron diffraction (RHEED) using a CCD camera. After growth, the samples are characterized by a DME DualScope 95-50 atomic force microscope (AFM), a Philips X'Pert MRD diffractometer for high resolution X-ray diffraction (HRXRD) (the resolution in $\omega$-direction is limited to about 6 arcsec by the instrumental broadening of the 4$\times$Ge(220) monochromator and the acceptance angle of the analyzer crystal is 12 arcsec) and a FEI Titan 80-300 scanning/transmission electron microscope (S/TEM) operated at 300kV. TEM cross-sectional specimens are prepared by focused ion beam milling.

	Previous experience has shown 300~$^\circ$C to be the optimal substrate temperature for growing Bi$_2$Se$_3$ in our MBE chamber. To reduce phosphorus out-diffusion from the InP substrate,\cite{Honig1969} we choose a two-temperature growth start (2T-start)\cite{Li2010,Bansal2011} to reach this temperature. We first heat the substrate to a temperature T$_s$~=~250~$^\circ$C and cover the InP with approximately two QLs of Bi$_2$Se$_3$. The sample is then heated to T$_s$~=~300~$^\circ$C and this initial Bi$_2$Se$_3$ layer is annealed in Se atmosphere to improve its crystal quality. This is followed by deposition of the bulk of the layer in normal MBE growth mode. As a control, we have also prepared a sample with a one-temperature growth start (1T-start), where we heat directly to T$_s$~=~300~$^\circ$C and start growth. For both methods, the Bi$_2$Se$_3$ layer is grown at a rate of approximately 1 nm per minute (0.17~\AA$s^{-1}$), using beam equivalent pressures (BEP) of p$_{Se}$~=~6~$\times$~10$^{-6}$ mbar and p$_{Bi}$~=~2~$\times$~10$^{-7}$ mbar. After growth, the samples are cooled to T$_s=140~^\circ$C in a Se atmosphere to avoid formation of Se vacancies. Below, structural data on four samples, A, B, C, and D of different Bi$_2$Se$_3$ layer thicknesses, corresponding to growth times (after anneal) of 22~min, 1~h, 3~h, and 6~h are presented for the 2T-start procedure. Sample E is grown by 1T-start procedure with a growth time of 1~h.   

\begin{figure}[b]                                
\begin{center}                                      
\includegraphics[width=1\linewidth]{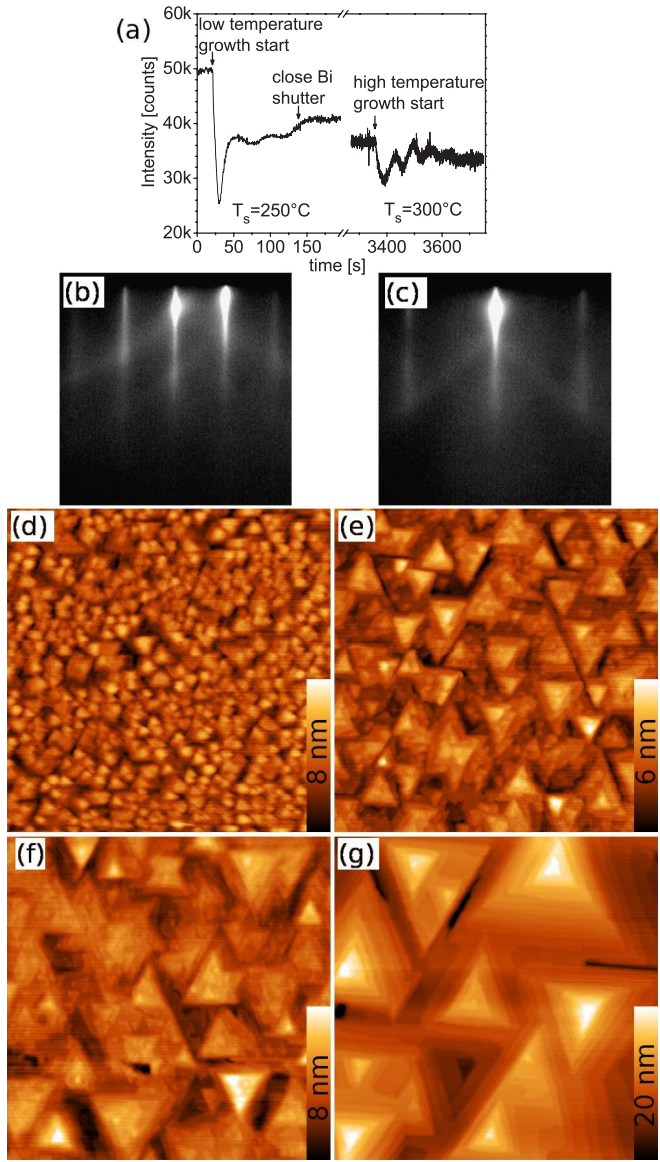}      
\caption[]{(a) RHEED specular spot intensity vs time of the Bi$_2$Se$_3$ growth start (T$_s$~=~250~$^\circ$C) as well as the beginning of the second growth step (T$_s$~=~300~$^\circ$C) for sample B, (b) and (c) RHEED images of the Bi$_2$Se$_3$ surface observed along the $[0\overline{1}1]$ and $[1\overline{2}1]$ directions of the substrate, after growth is completed, (d) - (g) AFM images (7~$\times$~7~$\mu$m, lower right corner: z scale bar labeled by the maximum height distribution in the image) of samples A, B, C, and D, respectively.}   
\label{1}                                      
\end{center}
\end{figure}

	RHEED is used to monitor the growth. Fig.~\ref{1}(a) shows the intensity of the specular spot as a function of time. We observe two oscillations assigned to the growth of about 2~QLs during the low temperature growth step. A slight increase of intensity continues when closing the Bi shutter, and saturates 25~s later. The sample is then annealed for $\sim$~1~h at T$_s$~=~300~$^\circ$C, and normal MBE growth is initiated. At this high-temperature growth start distinct oscillations are again observed. These gradually decrease in amplitude within the first 4~QLs. Extrapolating the layer thickness from the period of these oscillations and the growth time gives a film thickness of d$_A$~=~19~nm, d$_B$~=~54 nm, d$_C$~=~166~nm, d$_D$~=~315~nm, and d$_E$~=~53~nm for samples A through E, respectively. 
	
	RHEED streaks and Kikuchi lines\cite{Kainuma1955} are visible during the entire growth, and the FWHM of the streaks decrease with increasing layer thickness. The RHEED patterns in Figs.~\ref{1}(b) and \ref{1}(c) are obtained along the $[0\overline{1}1]$ and $[1\overline{2}1]$ directions of InP, respectively, after the second growth step of the Bi$_2$Se$_3$ layer of sample B. The change in streak separation between the $[0\overline{1}1]$ and $[1\overline{2}1]$ direction is a factor of $\sqrt{3}$, confirming the c-axis of the Bi$_2$Se$_3$ hexagonal unit cell is normal to the InP(111) planes.
	
	The AFM images of the samples (Figs.~\ref{1}(d)-(g)) reveal a surface comprised of domains of pyramidal shape, rotated by 60$^\circ$ with respect to each other, likely the result of twinning during heteroepitaxy. The valleys between these domains have a depth of a few tens of nm in the case of the 250~nm thick sample D. Distinct steps are observed on the flanks of the domains. Their height is $\sim$~1~nm, which corresponds to single QL steps. The average size of the triangles increases with layer thickness. This is presumably caused by domains of one orientation overgrowing domains of the other type, and thus, merging into one larger domain.

\begin{figure}[b]                               
\begin{center}                                      
\includegraphics[width=1\linewidth]{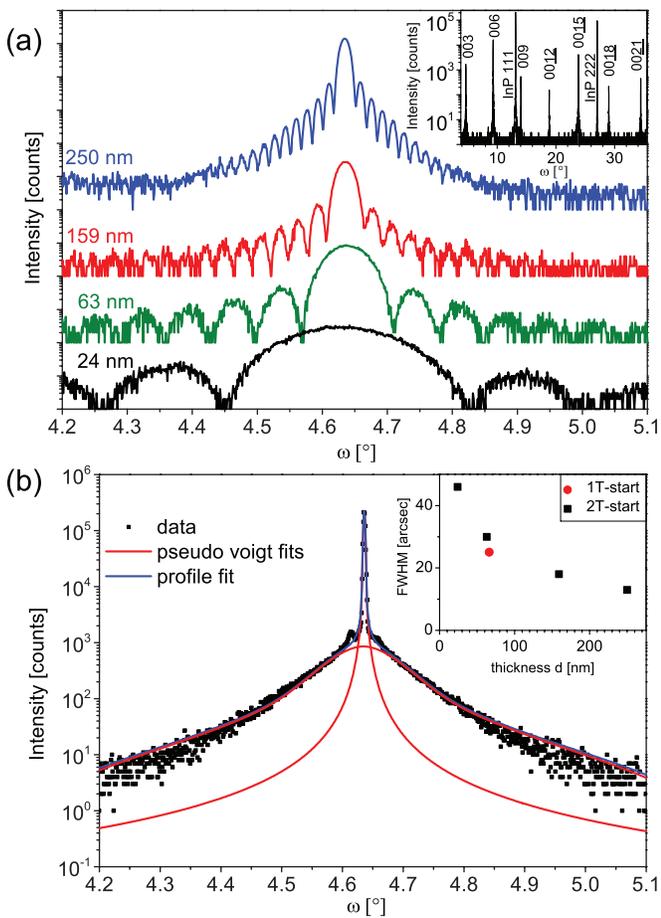}       
\caption[]{HRXRD of the Bi$_2$Se$_3$ (003) reflection on a logarithmic scale, (a) ($\omega-2\theta$) diffractograms for layers of various thicknesses, showing distinct thickness fringes (subsequent curves offset by two decades for clarity). The inset shows a large-scale ($\omega-2\theta$) scan with the peaks identified, (b) rocking curve scan of the 250~nm sample D with a FWHM of 13~arcsec, the profile fit (blue) is the superposition of two pseudo-Voigt fits (red), the inset shows the FWHM of rocking curves of the various samples plotted versus layer thickness.}
\label{2}                                    
\end{center}
\end{figure}
	
	The ($\omega-2\theta$) HRXRD scan of sample D presented in the inset of Fig.~\ref{2}(a) shows sharp and intense (00$l$) ($l=3n$, $n\in~\mathbb{N}_1$) peaks of Bi$_2$Se$_3$ in addition to the (111) and (222) peaks of the substrate. The absence of other peaks confirms the growth of primarily single stoichiometric phase Bi$_2$Se$_3$. The higher resolution ($\omega-2\theta$) scans of the (003) reflection (shown in Fig.~\ref{2}(a)) reveal pronounced layer thickness fringes. Their period corresponds to a layer thicknesses of d$_A$~=~24~nm, d$_B$~=~63~nm, d$_C$~=~159~nm, d$_D$~=~250~nm, and d$_E$~=~66~nm for samples A, B, C, D, and E, respectively. These are not inconsistent with the thicknesses estimated from RHEED when taking into account that the latter assumes a perfectly constant growth rate.
	
	A HRXRD rocking curve ($\omega$-scan) of sample D shows a very sharp (003) Bi$_2$Se$_3$ peak with a FWHM  $\Delta\omega_D$~=~13~arcsec (Fig.~\ref{2}(b)). This FWHM is comparable to the instrumental resolution. The inset of Fig.~\ref{2}(b) shows the measured FWHM $\Delta\omega$, which decreases with increasing layer thickness. This shows that the mosaicity of the layer decreases with thickness. The FWHM of samples E and B with similar layer thickness, but with different growth starts, is comparable. This suggests that not the annealing of the starting layer in the 2T-start, often used for improving the crystal quality on lattice mismatched substrates,\cite{Li2010,Bansal2011} but rather the InP(111) substrate results in the achieved high structural quality. The sharp (00~$l$) peaks are surrounded by a weak diffusely scattered background and shoulders in  $\omega$-direction. The data are fitted by two pseudo-Voigt profiles with 14\% integrated intensity for the background.

\begin{figure}[tb]                               
\begin{center}                                      
\includegraphics[width=1\linewidth]{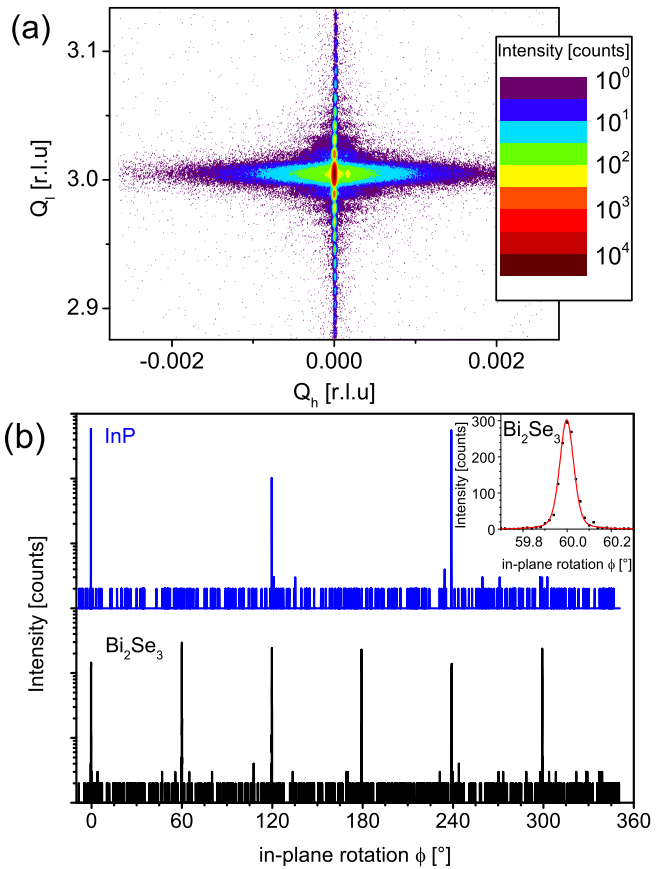}       
\caption[]{(a) Reciprocal space map of the Bi$_2$Se$_3$ (003) reflection of sample D. (b) Asymmetric reflections of \{0 1 5\} layer planes and  \{0 0 2\} substrate planes (curve offset by three decades for clarity) as a function of in-plane rotation $\phi$, the inset shows a close up of a layer peak.}
\label{3}                                    
\end{center}
\end{figure}

\begin{figure}[tb]                               
\begin{center}                                      
\includegraphics[width=1\linewidth]{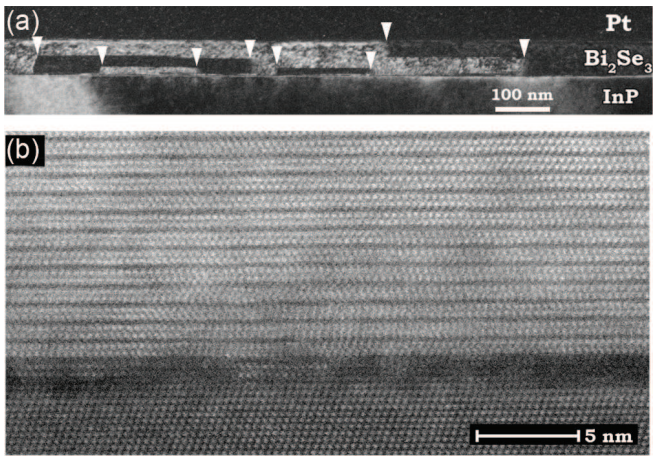}     
\caption[]{(a) Dark-field TEM image of sample B. White arrows indicate the positions of twin domain boundaries, (b) cross-sectional high-angle annular dark field S/TEM image of the Bi$_2$Se$_3$/InP interface region within a single domain.}
\label{4}
\end{center}
\end{figure}

	The reciprocal space map in Fig. \ref{3}(a) clearly shows the narrow (003) reflection of the 250 nm Bi$_2$Se$_3$ layer (sample D) along with the associated layer-thickness fringes and the shoulders, as well as a diffuse background feature, which is elongated in the Q$_{h}$ direction. The background and shoulders could be related to any combination of finite size effects\cite{Kidd1995} of the domains, tilts at domain boundaries, interface defects, or the slight ``waviness" of the QLs seen in TEM images (see Fig.~\ref{4}(b)). We note, however, that a direct comparison of our HRXRD data with the published data on epitaxial Bi$_2$Se$_3$ layers is difficult, as the reported $\omega$-widths may be increased by instrumental resolution.

	 Asymmetric reflections from the \{0 1 5\} Bi$_2$Se$_3$ layer planes and the \{0 0 2\} substrate planes are measured as a function of in-plane rotation $\phi$, in order to study twinning and the influence of the substrate's lattice on the layer. The Bi$_2$Se$_3$ reflections occur every 60$^\circ$ (see Fig. \ref{3}(b)), instead of the 120$^\circ$ expected for trigonal symmetry, consistent with reports of hot wall epitaxy of Bi$_2$Se$_3$ on InP(111)B.\cite{Takagaki2012} The observed six fold symmetry is caused by twin domains with either stacking order $ABC$ or the reversed $ACB$. The three \{0 0 2\} reflections of the substrate are at the same azimuthal angle $\phi$ as those of one triplet of the Bi$_2$Se$_3$ layer. This confirms that the hexagonal lattice of the Bi$_2$Se$_3$ layer is oriented parallel to that of the substrate, to within a twist of order 0.07$^\circ$, given by the FWHM of the Bi$_2$Se$_3$ reflections in $\phi$ direction (see inset Fig. \ref{3}(b)). Apparently, the van der Waals bonds between the Se and the substrate are sufficient to orient the hexagonal lattice structure, but do not determine the choice of ``$B$" or ``$C$" sites for the second atomic layer (Bi), allowing both $ABC$ and $ACB$ stacking, and thus the formation of twin domains.

	TEM studies of sample B further characterize layer and interface properties. Selected area electron diffraction patterns display sharp spots and confirm the parallel orientation of the Bi$_2$Se$_3$(001) and InP(111) planes. Both electron diffraction patterns and dark-field TEM images reveal the presence of rotational and lamellar twin domains in the layer (Fig.~\ref{4}(a)).\cite{Tarakina2012} The observed decreasing number of twin domain boundaries (indicated by white arrows) with distance to the interface supports the previously suggested mechanism of merging and overgrowth of twin domains, which results in a domain size increasing with layer thickness. At the bottom interface, there is a layer of relatively low crystalline quality along with higher crystalline quality regions up to a QL thickness, which is followed by well-crystallized QLs (Fig.~\ref{4}(b)).


In summary, using InP(111) substrates, which are nearly lattice-matched to the natural lattice constant of Bi$_2$Se$_3$ allows for the MBE growth of layers of high structural quality. HRXRD investigations reveal that the layers have low mosaicity-tilt, and -twist, and are of uniform thickness across the sample. AFM and TEM studies show that the layers are nevertheless not truly monocrystalline but rather comprised of pyramid shaped twin domains of a size which increases with layer thickness. The formation of twin domains appears to be an intrinsic property of heteroepitaxy of Bi$_2$Se$_3$ on hexagonal surfaces like InP(111) due to the two possible orders of layer stacking in the Bi$_2$Se$_3$ unit cell, i.e., $ABCABC$... and $ACBACB$...\cite{Tarakina2012} It would be of interest to study growth on InP substrates of various orientations to see if this property can be controlled. Finally, the length scale of the achieved domains is already sufficient to consider patterning mesoscopic devices such that they comprise of a single high quality domain.
\\
\begin{acknowledgments}
We wish to thank C. Ames for valuable discussions. We gratefully acknowledge the financial support of the EU ERC-AG Program (project 3-TOP). N.V.T. acknowledges funding by the Bavarian Ministry of Sciences, Research and the Arts. G.K. acknowledges funding by the Alexander von Humboldt Foundation.
\end{acknowledgments}
 
%

\end{document}